\documentclass[journal,draftcls,onecolumn,12pt,twoside]{IEEEtranTCOM}
\normalsize

\ifCLASSINFOpdf
\else
\fi

\usepackage[nospace]{cite}
\usepackage{latexsym, bm}
\usepackage{graphicx}
\usepackage{cite}
\usepackage{amssymb}
\usepackage{amsmath}
\usepackage{algorithm}
\usepackage[noend]{algorithmic}
\usepackage{subfigure}
\usepackage{enumerate}
\usepackage{makecell}

\usepackage{slashbox}
\usepackage{verbatim}
\begin{document}
%
\title{Cross-Layer Scheduling in Multi-User System with Delay and Secrecy Constraints}
%
%
%

\author{Jun Wang, Pengfei Huang, and Xudong Wang
\thanks{Jun Wang, Pengfei Huang, and Xudong Wang are with UM-SJTU Joint Institute,
Shanghai Jiao Tong University.
E-mail: jun.wangbreath@gmail.com, newday@sjtu.edu.cn, and wxudong@sjtu.edu.cn.}

}

\maketitle

\begin{abstract}

Recently, physical layer security based approaches have drawn considerable attentions and are envisaged to provide secure communications in the wireless networks.
However, most existing literatures only focus on the physical layer. Thus, how to design an effective transmission scheme which also considers the requirements from the upper layers is still an unsolved problem.
We consider such cross-layer resource allocation problem in the multi-user downlink environment for both having instantaneous and partial eavesdropping channel information scenarios.
The problem is first formulated in a new security framework. Then,
the control scheme is designed to maximize the average admission rate of the data, incorporating delay, power, and secrecy as constraints, for both non-colluding and colluding eavesdropping cases in each scenario.
Performance analysis is given based on the stochastic optimization theory and the simulations are carried out to validate the effectiveness of our scheme.

\section{Introduction}

Recently, physical layer security has drawn considerable interests, and is expected to provide secure communications in the wireless networks.
Physical layer security dates back to the Shannon's notion of perfect security \cite{shannon1949communication}, and then it is studied in \cite{wyner1975wire}  \cite{leung1978gaussian} \cite{csiszar1978broadcast}.
They show that secure communication is possible if the legitimate receiver has a better channel than the eavesdropper.
The impact of channel fading is lately considered very helpful that perfect secrecy can be achieved even when the eavesdropping channel is stronger than the legitimate channel on average \cite{bloch2006wireless} \cite{gopala2008secrecy}\cite{liang2008secure}.

So far many progresses are made to enhance the secrecy with the advanced physical layer technologies.
One of these is to employ beamforming to strengthen the quality of the legitimate link in the
multi-antenna systems. In \cite{shafiee2007achievable}, beamforming is proved to be the optimal strategy for secrecy in the MISO system. Then, the robust power allocation to maximize the secrecy rate for the MISO system is studied in \cite{huang2011robust} and  \cite{li2011optimal}. To further improve the secrecy, artificial noise is used to degrade the eavesdropping channel \cite{goel2008guaranteeing}.
 Based on \cite{goel2008guaranteeing}, beamforming and artificial noise are shown to evidently improve secrecy in the MIMO-OFDM system \cite{romero2011physical}.
 In \cite{6094170}, an optimization problem is investigated, which aims to minimize the total power consumption on both data and artificial noise to satisfy the minimum SNR at the legitimate user and a given average SNR at each eavesdropper.
 In \cite{zhou2010secure}, an analytical closed-form of the ergodic secrecy capacity of a single legitimate link in the presence of some eavesdroppers is calculated, and then the optimal power allocation between the data and artificial noise is also derived.
More recently, in contrast to the secrecy outage formulated in \cite{bloch2006wireless}, a new formulation which can depict reliability and security separately is proposed in \cite{zhou2011rethinking}.
Under this new framework, the benefits of the multiple transmitting antennas are investigated in \cite{zhangbenefits}.

 However, most efforts are made only in the physical layer. Thus, the interaction between the secrecy requirement in the physical layer and other QoS requirements (e.g., delay) in the upper layers of the wireless
  networks has not been sufficiently understood.
 So far a few papers have been published to solve this problem under the stochastic optimization framework (The stochastic optimization tool is used widely as in \cite{eryilmaz2006joint}, \cite{neely2006energy}, and \cite{neely2010stochastic}).
In \cite{koksal2010control}, a single hop uplink scenario is considered, where each node is controlled to send messages securely from other nodes with the objective of maximizing an overall utility.
In \cite{mao2011towards}, under a point to point secure communication scenario, the scheduling of the data, which is protected by either the physical layer security coding or the secret key, is investigated.
In \cite{liang2008wireless}, the broadcast channel model is considered, and the arrival rate supported by the
fading wiretap channel is analyzed and the power allocation policy is derived.

In this paper, we consider a different problem from above papers.
First, we focus on the multi-user downlink scenario like \cite{ioannis2011feedback}, and adopt a new security framework which can describe reliability and security separately, thus providing insight on the cross-layer resource allocation problem.
Second, we adopt beamforming and artificial noise as the physical layer technique. Then, a cross-layer power control scheme is carefully designed for both the total power allocation and the power ratio between the data and artificial noise, jointly considering delay, secrecy, energy consumption and multiuser diversity.
Third, we focus on two scenarios. One is the sender has instantaneous eavesdropping channel information. The other one is that the sender only has partial eavesdropping channel information.
In each scenario, both non-colluding and colluding eavesdropping cases are discussed.

The remainder of this paper is organized as follows. The system model is given in Section II. The optimization problem is formulated in Section III. The control scheme and the performance analysis are presented in Section IV. Simulation results are given in Section V and the paper is concluded in Section VI.

\section{System Model}

  We consider the secure communication between a transmitter (Alice) and $K$ legitimate receivers (i.e., $K$ Bobs) in the presence of $N_E$ eavesdroppers (Eves), as shown in Fig. \ref{Model}.
  Like \cite{ioannis2011feedback}, the transmitter Alice is equipped with $N_A$ antennas and each legitimate receiver Bob has one antenna.
  The time is considered slotted. At each time slot, the transmitter sends information to a single receiver based on the time division multiple access (TDMA) scheme.
  In addition, each Eve is equipped with a single antenna. Non-colluding and colluding cases are considered.
  In the former case, each Eve individually decodes the intercepted information.
  While in the later case, $N_E$ Eves jointly process their received information and we assume $N_A > N_E$ as same as in \cite{zhou2010secure}.

  We assume all the wireless links experience Rayleigh block fading. The channel gain varies from one time slot to another independently. In each time slot, the channel gain remains stable.
 During time slot $t$,
 we define $\textbf{h}_i(t)$ as the $1\times N_{A}$ channel gain vector between Alice and Bob $i$ ($i\in\{1,\ldots,K\}$), and its element is distributed as $\mathcal{CN}(0,1)$.
  $\textbf{g}_j(t)$ is the $1\times N_{A}$ channel gain vector between Alice and Eve $j$ ($j\in\{1,\ldots,N_E\}$), and its element is distributed as $\mathcal{CN}(0,1)$.
  $\textbf{G}(t)$ is the $N_{E}\times N_{A}$ channel gain matrix between Alice and colluding $N_E$ eavesdroppers. Each element is distributed as $\mathcal{CN}(0,1)$.
  $w$ is additive white Gaussian noise with distribution $\mathcal{CN}(0,\sigma_w^2)$. $\textbf{w}$ represents $N_E\times 1$ additive white Gaussian noise vector at $N_E$ colluding Eves and its distribution is $\mathcal{CN}(0, \textbf{I}\sigma_w^2)$.
  Without loss of generality, the noise is normalized with unit variance ($\sigma_w^2=1$).

  We assume Alice can accurately obtain the instantaneous channel information between Alice and all $K$ legitimate users.
  However, for the eavesdroppers,
  we consider two scenarios.
  The first scenario is that Alice can obtain the instantaneous eavesdropping channel information.
  The second scenario is that Alice only has partial information about the eavesdropping channel. More specifically,
  we assume Alice knows the number of Eves and the eavesdropping channel exhibiting Rayleigh fading, but Alice can not obtain the instantaneous eavesdropping channel gains.
  In each scenario, both non-colluding and colluding cases are considered.

\begin{figure}[!t]
\centering
\includegraphics[width=8cm]{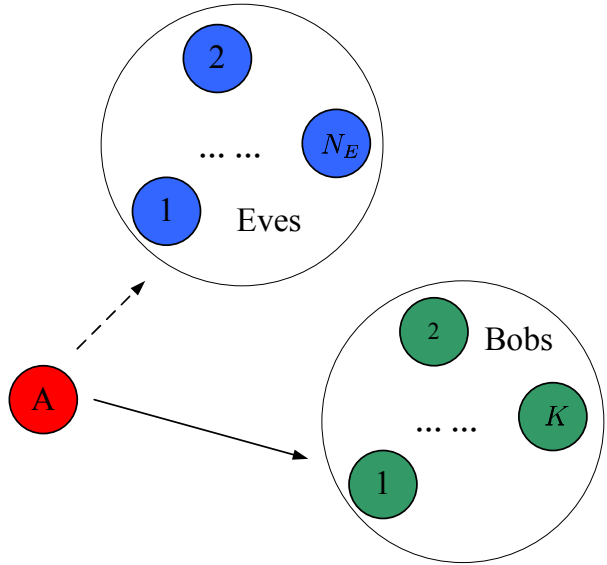}
\caption{System model} \label{Model}
\end{figure}

\section{Problem Formulation}
In this section, a cross-layer problem is formulated.
In the physical layer, the beamforming and the artificial noise are used to secure the data. While in the upper layer, the data queue of each user is required to be stable.
We jointly consider the requirements from different layers as follows.

\subsection{Channel Capacity with Beamforming and Artificial Noise}

 Beamforming and artificial noise are used as the physical layer technique to improve secrecy \cite{goel2008guaranteeing} \cite{zhou2010secure}, and it is described as follows. At time slot $t$, Alice generates an $N_{A}\times N_{A}$ matrix $\textbf{Z}(t)=[\textbf{z}_{1}(t) \ \textbf{Z}_{2}(t)]$, where $\textbf{z}_{1}(t)=\frac{\textbf{h}^*(t)}{||\textbf{h}(t)||}$ and $\textbf{Z}_{2}(t)$ is the null space matrix of \textbf{h}(t).
  The $N_{A} \times 1$ transmitted symbol vector by Alice is given as
  $\textbf{x}(t)=\textbf{z}_{1}(t)u(t)+\textbf{Z}_{2}(t) \textbf{v}(t)$.
  The variance of $u(t)$ is $\sigma_{u}^2(t)$ and each element of the $(N_{A}-1)\times 1$ vector
  $\textbf{v}(t)$ has circular symmetric complex Gaussian distribution with variance $\sigma_{v}^2(t)$.
  $u(t)$ and $\textbf{v}(t)$ represent data and artificial noise, respectively.
  The total power for the data and artificial noise is $P(t)$. Thus, $P(t)=\sigma_{u}^2(t)+(N_{A}-1)\sigma_{v}^2(t)$.
  We denote the fraction of the total power allocated to the data is $\varepsilon(t)$. Therefore, $\sigma_{u}^2(t)=\varepsilon(t) P(t)$ and $\sigma_{v}^2(t)=\frac{(1-\varepsilon(t)) P(t)}{N_A-1}$.

The legitimate channel between Alice and Bob $i$ is
\begin{equation} \label{eq:1}
\begin{split}
y_{bi}(t)&=\textbf{h}_i(t)\textbf{x}(t)+w   \\
         &=\textbf{h}_i(t)\textbf{z}_{1}(t)u(t)+\textbf{h}_i(t)\textbf{Z}_{2}(t)\textbf{v}(t)+w\\
         &=||\textbf{h}_i(t)||^2u(t)+w.
\end{split}
\end{equation}
The corresponding capacity of the legitimate channel between Alice and Bob $i$ is a function of
the control parameters $P(t)$ and $\varepsilon(t)$, and is denoted as $C_{bi}(t, P(t), \varepsilon(t))$
\begin{equation} \label{eq:2}
\begin{split}
C_{bi}(t, P(t), \varepsilon(t))&=log_{2}(1+\sigma_{u}^2(t)||\textbf{h}_{i}(t)||^2/\sigma^2_w)\\
&=log_{2}(1+\varepsilon(t) P(t)||\textbf{h}_{i}(t)||^2).
\end{split}
\end{equation}

In the non-colluding case, the eavesdropping channel between Alice and Eve $j$ is modeled as
\begin{equation} \label{eq:3}
\begin{split}
y_{ej}(t) &=\textbf{g}_j(t)\textbf{x}(t)+w \\
         &=\textbf{g}_j(t)\textbf{z}_{1}(t)u(t)+\textbf{g}_j(t)\textbf{Z}_{2}(t)\textbf{v}(t)+w.
\end{split}
\end{equation}
The corresponding capacity of the eavesdropping channel between Alice and Eve $j$ is denoted as $C_{ej}(t, P(t), \varepsilon(t))$
\begin{equation} \label{eq:4}
\begin{split}
&C_{ej}(t, P(t), \varepsilon(t))\\
=&log_{2}(1+\frac{|\textbf{g}_j(t)\textbf{z}_{1}(t)|^2\sigma^2_{u}(t)}{\big(\textbf{g}_j(t)\textbf{Z}_{2}(t)
\textbf{Z}^*_{2}(t)\textbf{g}^*_j(t)\big)\sigma^2_{v}(t)+\sigma^2_{w}  }).  \\
\end{split}
\end{equation}

Similarly, in the colluding case, the eavesdropping channel between Alice and colluding Eves is
\begin{equation} \label{eq:5}
\begin{split}
\textbf{y}_{eves}(t)=&\textbf{G}(t)\textbf{x}(t)+\textbf{w}\\
     =&\textbf{G}(t)\textbf{z}_{1}(t)u(t)+\textbf{G}(t)\textbf{Z}_{2}(t)\textbf{v}(t)+\textbf{w}.\\
    =&\bar{\textbf{g}}_{1}(t)u(t)+\bar{\textbf{G}}_{2}(t)\textbf{v}(t)+\textbf{w},
\end{split}
\end{equation}
where $\bar{\textbf{g}}_{1}(t)=\textbf{G}(t)\textbf{z}_{1}$(t) and $\bar{\textbf{G}}_{2}(t)=\textbf{G}(t)\textbf{Z}_{2}(t)$.
The corresponding capacity of the eavesdropping channel between Alice and the colluding Eves is denoted as $C_{eves}(t, P(t), \varepsilon(t))$
\begin{equation} \label{eq:6}
\begin{split}
&C_{eves}(t, P(t), \varepsilon(t))\\
=&log_{2}(\frac{| \bar{\textbf{G}}_{2}(t)  \bar{\textbf{G}}^*_{2}(t)    \sigma^2_{v}(t)+\textbf{I}\sigma^2_{w}+
\bar{\textbf{g}}_{1}(t) \bar{\textbf{g}}^*_{1}(t)  \sigma^2_{u}(t) |}
{|  \bar{\textbf{G}}_{2}(t)        \bar{\textbf{G}}^*_{2}(t)    \sigma^2_{v}(t)+\textbf{I}\sigma^2_{w}  |}     ) .
\end{split}
\end{equation}

\subsection{New Formulation of the Secrecy}

We consider a new security framework which can depict reliability and security separately as proposed in \cite{zhou2011rethinking}.
For the secure communication between Alice and Bob $i$, Alice chooses two rates. The rate of the transmitted codewords $R_{bi}(t)$ and the rate of the confidential information $R_{si}(t)$. $R_e(t)=R_{bi}(t)-R_{si}(t)$ reflects the cost of securing the messages against the eavesdropping. For each transmission, Bob $i$ can decode correctly if $C_{bi}(t, P(t), \varepsilon(t))>R_{bi}(t)$. While perfect secrecy fails if the eavesdropping channel capacity $C_e(t)$ is larger than $R_e(t)$. The secrecy outage probability $\mathrm{P_{so}}$ is defined as in \cite{zhou2011rethinking}
\begin{equation} \label{eq:7}
\mathrm{P_{so}}=\mathbb{P}\big(C_e(t)>R_e(t)|\text{message transmission}\big).
\end{equation}
Thus, the reliability ($R_{bi}(t)$) and security ($R_e(t)$) can be considered in (\ref{eq:2}) and (\ref{eq:7}), separately.

\subsubsection{Instantaneous Eavesdropping Channel Information Scenario}
 We assume we can obtain the instantaneous eavesdropping channel information of all $N_E$ eavesdroppers. Thus, we can achieve perfect secrecy, i.e., $\mathrm{P_{so}}=0$. For the secure communication between Alice and Bob $i$, the secrecy rate at time slot $t$ is a function of the control parameters $P(t)$ and $\varepsilon(t)$, denoted as $R_{si}(t, P(t), \varepsilon(t))$. For both non-colluding and colluding cases, $R_{si}(t, P(t), \varepsilon(t))$ can be calculated as follows
\begin{itemize}
  \item Non-colluding case
\begin{equation} \label{eq:8}
\begin{split}
R_{si}(t, P(t), \varepsilon(t))=&[R_{bi}(t)-R_e(t)]^+ \\
       =&[C_{bi}(t, P(t), \varepsilon(t))-\max\limits_{j\in\left\{1,2,\ldots, N_E\right\}} C_{ej}(t, P(t), \varepsilon(t))]^+,
\end{split}
\end{equation}
where $[a]^+$ is $\text{max}[a,0]$.
  \item Colluding case
  \begin{equation} \label{eq:9}
\begin{split}
R_{si}(t, P(t), \varepsilon(t))=&[R_{bi}(t)-R_{e}(t)]^+ \\
       =&[C_{bi}(t, P(t), \varepsilon(t))-C_{eves}(t, P(t), \varepsilon(t))]^+.
\end{split}
\end{equation}

\end{itemize}

\subsubsection{Partial Eavesdropping Channel Information Scenario}
Since we can not obtain the instantaneous eavesdropping channel state,
it results in:
1) whether message is transmitted is independent from the current eavesdropping channel state, i.e.,
(\ref{eq:7}) is converted into the unconditional probability: $\mathrm{P_{so}}=\mathbb{P}\big(C_e(t)>R_e(t)\big)$;
2) the perfect secrecy can not be guaranteed.
Thus, we focus on designing the transmission scheme such that the secrecy outage $\mathrm{P_{so}}$ can satisfy certain secure level $\eta$.
For the communication between Alice and Bob $i$, the secrecy rate $R_{si}(t, P(t), \varepsilon(t), \eta)$ at time slot $t$ for non-colluding and colluding cases are derived as follows.

\begin{itemize}
  \item Non-colluding case

For $N_E$ non-colluding eavesdroppers, the secrecy outage $\mathrm{P_{so}}$ is expressed as $1-[\mathbb{P}(C_{ej}(t, P(t), \varepsilon(t))<R_e(t))]^{N_E}$.
Since the detailed distribution of $C_{ej}(t, P(t), \varepsilon(t))$ is complex,
we consider the worst case that the SNR at the eavesdropper is very high so that $\sigma^2_{w}$ is negligible compared to the
artificial noise. By omitting $\sigma^2_{w}$ in (\ref{eq:4}), we
obtain the upper bound of $C_{ej}(t, P(t), \varepsilon(t))$, denoted as $C_{ej}^{up}(t, \varepsilon(t))$,
\begin{equation} \label{eq:10}
\begin{split}
&C_{ej}^{up}(t, \varepsilon(t))\\
=& log_{2}(1+\frac{|\textbf{g}_j(t)\textbf{z}_{1}(t)|^2\sigma^2_{u}(t)}{\big(\textbf{g}_j(t)\textbf{Z}_{2}(t)
\textbf{Z}^*_{2}(t)\textbf{g}^*_j(t)\big)\sigma^2_{v}(t)})\\
=& log_{2}(1+ \frac{|\textbf{g}_j(t)\textbf{z}_{1}(t)|^2 (N_A-1)\varepsilon(t)}{\big(\textbf{g}_j(t)\textbf{Z}_{2}(t)
\textbf{Z}^*_{2}(t)\textbf{g}^*_j(t)\big)(1-\varepsilon(t))}),
\end{split}
\end{equation}
where
$\frac{|\textbf{g}_j(t)\textbf{z}_{1}(t)|^2 (N_A-1) } {\textbf{g}_j(t)\textbf{Z}_{2}(t)
\textbf{Z}^*_{2}(t)\textbf{g}^*_j(t)}$ has a distribution as F-distribution with parameter $(2,2N_A-2)$, and its probability
density function is
$f(x)=\frac{(N_A-1)^{N_A}}{(x+N_A-1)^{N_A}}$ \cite{zhou2010secure}.

Thus, to ensure that $\mathrm{P_{so}}$ satisfy secure level $\eta$, we let $\mathrm{P_{so}^{up}}$ (i.e., $1- \big[ \mathbb{P}\big(C_{ej}^{up}(t, \varepsilon(t))<R_e(t)\big) \big]^{N_E}$) satisfy the secrecy outage requirement, where $\mathbb{P}\big(C_{ej}^{up}(t, \varepsilon(t))<R_e(t)\big)$ is calculated as follows
\begin{equation} \label{eq:11}
\begin{split}
&\mathbb{P}\big(C_{ej}^{up}(t, \varepsilon(t))<R_e(t)\big)\\
=&\mathbb{P}\big(log_2(1+\frac{\varepsilon(t)}{1-\varepsilon(t)}x)<R_e(t)\big)\\
=&\mathbb{P}\big(x<(2^{R_e(t)}-1)\frac{1-\varepsilon(t)}{\varepsilon(t)}\big)\\
=&-(N_A-1)^{N_A-1}(x+N_A-1)^{1-N_A}+1|_{x=(2^{R_e(t)}-1)\frac{1-\varepsilon(t)}{\varepsilon(t)}}\\
=&-(N_A-1)^{N_A-1}((2^{R_e(t)}-1)\frac{1-\varepsilon(t)}{\varepsilon(t)}+N_A-1)^{1-N_A}+1.
\end{split}
\end{equation}
For secure level $\eta$, let $\mathrm{P_{so}^{up}}=\eta$, so the relationship between $R_e(t)$ and $\varepsilon(t)$ is
\begin{equation} \label{eq:12}
\begin{split}
&-(N_A-1)^{N_A-1}((2^{R_e(t)}-1)\frac{1-\varepsilon(t)}{\varepsilon(t)}+N_A-1)^{1-N_A}\\
=&(1-\eta)^{\frac{1}{N_E}}-1.
\end{split}
\end{equation}
When $N_A=6$ and $N_E=3$, the relationship between $R_e(t)$ and $\varepsilon(t)$ is shown in Fig. \ref{noncolludingoutageprobability}.
Since $R_e(t)$ is determined by $\varepsilon(t)$ and $\eta$, $R_e(t)$ is denoted as $R_e(t, \varepsilon(t), \eta)$. Thus, the secrecy rate $R_{si}(t, P(t), \varepsilon(t), \eta)$ is determined as follows.
\begin{equation} \label{eq:13}
\begin{split}
&R_{si}(t, P(t), \varepsilon(t), \eta) \\
=&[R_{bi}(t)-R_e(t)]^+ \\
       =&[C_{bi}(t, P(t), \varepsilon(t))-R_e(t, \varepsilon(t), \eta)]^+, \\
\end{split}
\end{equation}
where $C_{bi}(t, P(t), \varepsilon(t))$ is calculated by (\ref{eq:2}), and $R_e(t, \varepsilon(t), \eta)$ is computed by (\ref{eq:12}).

\begin{figure}[!t]
\centering
\includegraphics[width=8cm]{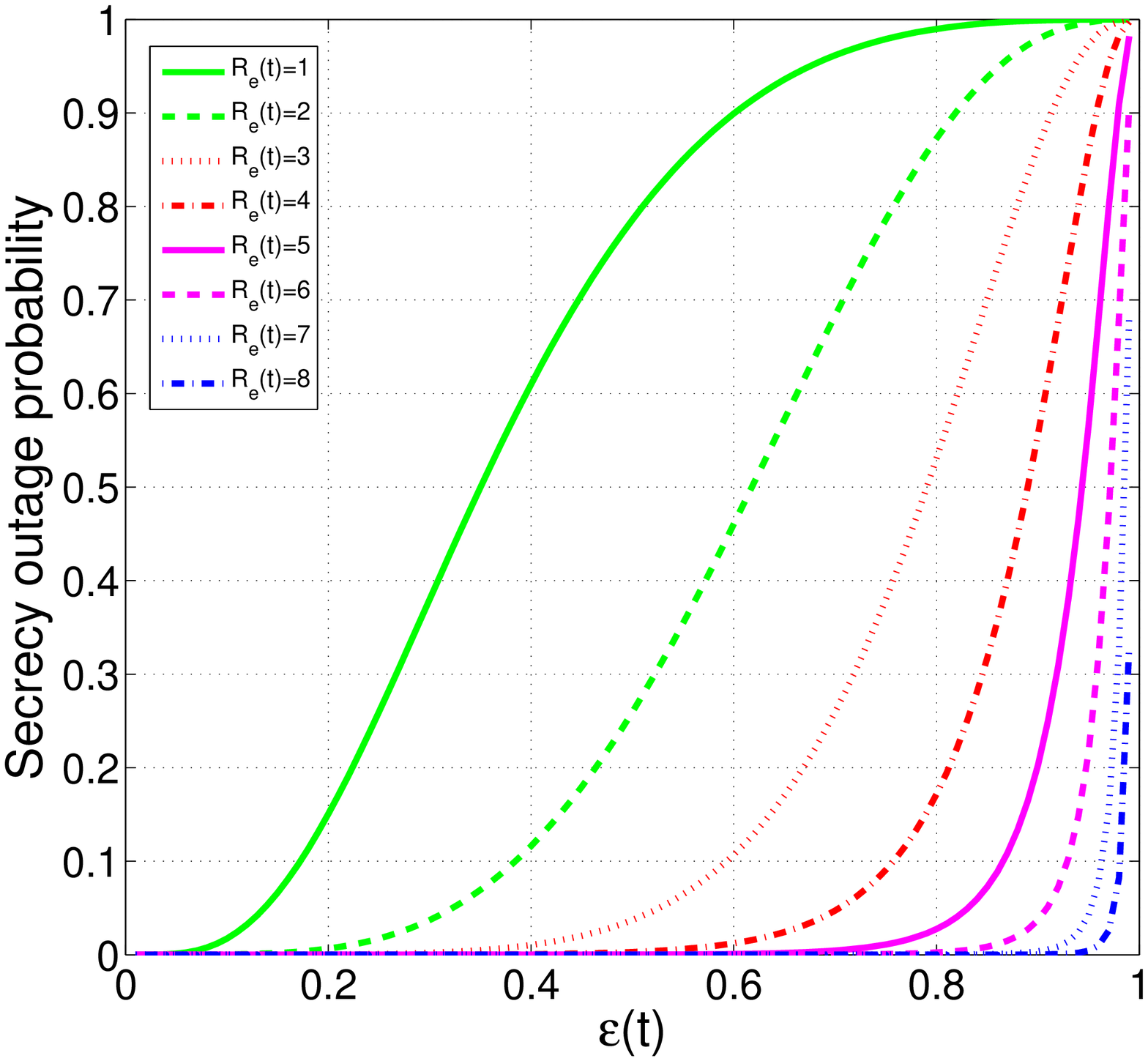}
\caption{Secrecy outage probability versus $\varepsilon(t)$ for the non-colluding case in partial eavesdropping channel information scenario when $N_A=6$ and $N_E=3$. } \label{noncolludingoutageprobability}
\end{figure}

  \item Colluding case

Similar to the non-colluding case, for $N_E$ colluding Eves, the secrecy outage $\mathrm{P_{so}}$ can be expressed as $\mathbb{P}\big(C_{eves}(t, P(t), \varepsilon(t))>R_e(t)\big)$. We obtain the upper bound of $C_{eves}(t, P(t), \varepsilon(t))$, denoted as $C_{eves}^{up}(t, \varepsilon(t))$, by omitting $\sigma^2_{w}$ in (\ref{eq:6}),
\begin{equation} \label{eq:14}
\begin{split}
&C_{eves}^{up}(t, \varepsilon(t))\\
=&log_{2}\big|\textbf{I}+\sigma_u^2(t)\bar{\textbf{g}}_{1}(t)\bar{\textbf{g}}_{1}^*(t)\big(\sigma_v^2(t)\bar{\textbf{G}}_{2}(t)\bar{\textbf{G}}_{2}^*(t) \big)^{-1} \big|\\
=&log_{2}\big(1+  \frac{N_A-1}{\varepsilon(t)^{-1}-1}\bar{\textbf{g}}_{1}^*(t)\big(\bar{\textbf{G}}_{2}(t)\bar{\textbf{G}}_{2}^*(t)  \big)^{-1}\bar{\textbf{g}}_{1}(t)\big),\\
\end{split}
\end{equation}
where $\bar{\textbf{g}}_{1}^*(t)(\bar{\textbf{G}}_{2}(t)\bar{\textbf{G}}_{2}^*(t) )^{-1}\bar{\textbf{g}}_{1}(t)$ has a distribution that its complementary cumulative distribution function is
$F^c(x)=\frac{\sum_{k=0}^{N_E-1} {N_A-1 \choose k}x^k }{(1+x)^{N_A-1}}$ \cite{zhou2010secure}.

Thus, to ensure that $\mathrm{P_{so}}$ satisfy secure level $\eta$, we let $\mathrm{P_{so}^{up}}$ \big(i.e., $\mathbb{P}\big(C_{eves}^{up}(t, \varepsilon(t))>R_e(t)\big)$\big) satisfy this secrecy outage requirement,
\begin{equation} \label{eq:15}
\begin{split}
\mathrm{P_{so}^{up}}=&\mathbb{P}(C_{eves}^{up}(t)>R_e(t))\\
=&\mathbb{P}(C_{eves}^{up}(t)>R_e(t))\\
=&\mathbb{P}(log_2(1+\frac{N_A-1}{\varepsilon(t)^{-1}-1}x)>R_e(t))\\
=&\mathbb{P}(x>(2^{R_e(t)}-1)\frac{\varepsilon(t)^{-1}-1}{N_A-1})\\
=&\frac{\sum_{k=0}^{N_E-1} {N_A-1 \choose k}x^k }{(1+x)^{N_A-1}}|_{x=(2^{R_e(t)}-1)\frac{\varepsilon(t)^{-1}-1}{N_A-1}}\\
=&\frac{\sum_{k=0}^{N_E-1} {N_A-1 \choose k}x^k }{(1+(2^{R_e(t)}-1)\frac{\varepsilon(t)^{-1}-1}{N_A-1})^{N_A-1}}.
\end{split}
\end{equation}
For secure level $\eta$, let $\mathrm{P_{so}^{up}}=\eta$, so the relationship between $R_e(t)$ and $\varepsilon(t)$ is
\begin{equation} \label{eq:16}
\begin{split}
\frac{\sum_{k=0}^{N_E-1} {N_A-1 \choose k}x^k }{(1+(2^{R_e(t)}-1)\frac{\varepsilon(t)^{-1}-1}{N_A-1})^{N_A-1}}=\eta.
\end{split}
\end{equation}
For $N_A=6$ and $N_E=3$, the relationship between $R_e(t)$ and $\varepsilon(t)$ is shown in Fig. \ref{colludingoutageprobability}.
Since $R_e(t)$ is determined by $\varepsilon(t)$ and $\eta$, $R_e(t)$ is denoted as $R_e(t, \varepsilon(t), \eta)$. Thus, the secrecy rate $R_{si}(t, P(t), \varepsilon(t), \eta)$ is determined as follows
\begin{equation} \label{eq:17}
\begin{split}
R_{si}(t, P(t), \varepsilon(t), \eta)=&[R_{bi}(t)-R_e(t)]^+ \\
       =&[C_{bi}(t, P(t), \varepsilon(t))-R_e(t, \varepsilon(t), \eta)]^+, \\
\end{split}
\end{equation}
where $C_{bi}(t, P(t), \varepsilon(t))$ is calculated by (\ref{eq:2}), and $R_e(t, \varepsilon(t), \eta)$ is computed by (\ref{eq:16}).

\begin{figure}[!t]
\centering
\includegraphics[width=8cm]{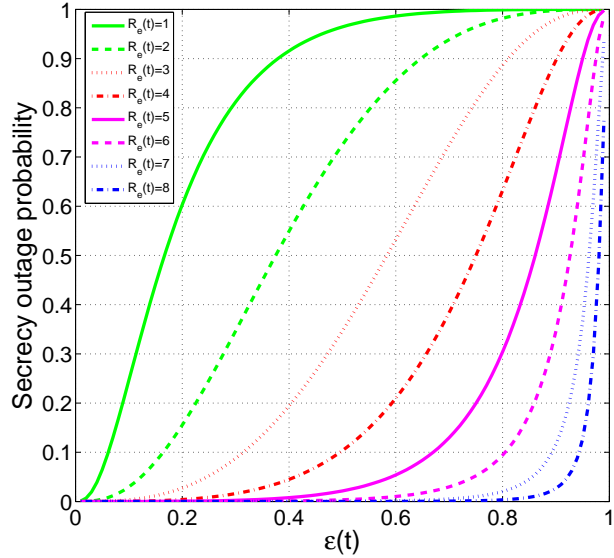}
\caption{ Secrecy outage probability versus $\varepsilon(t)$ for the colluding case in partial eavesdropping channel information scenario when $N_A=6$ and $N_E=3$.} \label{colludingoutageprobability}
\end{figure}

\end{itemize}

In the partial eavesdropping channel information scenario,
for both non-colluding and colluding cases,
when $\varepsilon(t)=0$, i.e., no data is transmitted, $R_e(t, \varepsilon(t), \eta)$ is defined as $0$ for any $\eta$. Thus, $R_{si}(t, P(t), \varepsilon(t), \eta)=0$.
When $\varepsilon(t)=1$, i.e., no artificial noise is generated, $R_e(t, \varepsilon(t), \eta)$ is defined as $+\infty$ for any $\eta$. Thus, $R_{si}(t, P(t), \varepsilon(t), \eta)=0$.

\subsection{Upper Layer Data Queue Process}
In the upper layer, the data queue process is considered.
For user $i$ ($i=1,\ldots,K$) at time slot $t$, let $A_i(t)$ denote the data arrival process, and it is bounded by $A_{max}$. Only $R_i(t)$ of $A_i(t)$ are admitted into the data queue $U_i(t)$ in order to keep the data queue stable.
At time slot $t$, only one user $i$ is served with rate $R_{si}(t, P(t), \varepsilon(t), \eta)$, which is related to the secrecy outage requirement $\eta$, the total power $P(t)$ and power ratio $\varepsilon(t)$.
Data queue $U_i(t)$ is updated as follows
\begin{equation}  \label{eq:18}
\begin{split}
U_i(t+1)=\text{max}[U_i(t)-R_{si}(t, P(t), \varepsilon(t), \eta)I_i(t),0 ]+R_i(t),
\end{split}
\end{equation}
where $I_i(t)$ is an indicator. If $I_i(t)=1$, it means user $i$ is chosen for transmission at time slot $t$.

\subsection{Optimization Problem Formualtion} \label{formulation}
We consider the optimization problem incorporating both the physical layer and upper layer. Let $r_i$ denote the long term average admission rate of the data for user $i$, i.e., $r_i=\lim _{T\rightarrow\infty}\frac{1}{T}\sum_{t=0}^{T-1} \mathbb{E}\{R_i(t)\} $. Let $\{\theta_1, \theta_2, \ldots, \theta_K\}$ be a collection of positive weights. Our objective is to maximize the sum of weighted average admission rate under the average power, secrecy and queue stability constraints. The optimization problem can be formulated as follows
\begin{displaymath}
\begin{split}
\text{Maximize}  \sum_{i=1}^{K}\theta_i r_i
\end{split}
\end{displaymath}
s.t.
\begin{eqnarray}
&\displaystyle{\limsup_{T\rightarrow\infty}\frac{1}{T}\sum_{t=0}^{T-1}U_i(t) \leq +\infty}, i=1,\ldots,K, \label{eq:19} \\
&\displaystyle{\limsup_{T\rightarrow\infty}\frac{1}{T}\sum_{t=0}^{T-1}P(t) \leq  P_{av}}, \label{eq:20}       \\
&0\leq R_i(t)\leq A_i(t), i=1,\ldots,K, \label{eq:21}\\
&P(t)\in \Pi, \varepsilon(t) \in \Lambda, \mathrm{P_{so}}\leq \eta \label{eq:22}.
\end{eqnarray}
 In the above constraints, (\ref{eq:19}) requires the data queue to be stable. (\ref{eq:20}) describes the average power constraint. (\ref{eq:21}) shows that the admitted data is less than the arrival data. (\ref{eq:22}) means the selection set for total power $P(t)$ is $\Pi$, the selection set for power ratio $\varepsilon(t)$ is $\Lambda$, and the secrecy outage constraint is $\eta$.

Similar to \cite{neely2006energy} \cite{neely2010stochastic}, to satisfy the average power constraint, a virtual power queue $X(t)$ is defined. It is updated as follows
\begin{equation}   \label{eq:23}
\begin{split}
X(t+1)=\text{max}[X(t)-P_{av},0]+P(t).
\end{split}
\end{equation}
\section{Control Scheme and Performance Analysis}
In this section, we present the control scheme and also give the performance analysis, using the stochastic optimization tool.

\subsection{Cross-layer Control Scheme}

1. Admission control:

\begin{displaymath}
\begin{split}
\text{Minimize}  & \sum_{i=1}^K   (U_i(t)-V\theta_i) R_i(t), \\
&0\leq R_i(t)\leq A_i(t).\\
\end{split}
\end{displaymath}

2. Power allocation:
\begin{displaymath}
\begin{split}
\text{Maximize} & \sum_{i=1}^K U_i(t) R_{si}(t, P(t), \varepsilon(t), \eta)I_i(t) -  X(t)P(t), \\
&P(t)\in \Pi, \varepsilon(t) \in \Lambda, \mathrm{P_{so}}\leq \eta,\\
\end{split}
\end{displaymath}
where the secrecy rate $R_{si}(t, P(t), \varepsilon(t), \eta)$ is calculated differently under different conditions.
When in the scenario that the instantaneous eavesdropping channel information is available, perfect secrecy (i.e., $\eta=0$) can be achieved. $R_{si}(t, P(t), \varepsilon(t), \eta)$ are calculated according to (\ref{eq:8}) and (\ref{eq:9}) for non-colluding and colluding cases, respectively.
When in the scenario that only partial eavesdropping channel information is available, a non-zero secrecy outage requirement $\eta$ can be satisfied. $R_{si}(t, P(t), \varepsilon(t), \eta)$ are calculated according to (\ref{eq:13}) and (\ref{eq:17}) for non-colluding and colluding cases, respectively.

3. Queue update:

The data queue $U_i(t)$ and the virtual power queue $X(t)$ are updated according to (\ref{eq:18}) and (\ref{eq:23}), respectively.

Control scheme proof:

The proof is similar to \cite{neely2006energy} \cite{neely2010stochastic}.
We define $Q(t)$ as a vector of all queues $ Q(t)=(U_1(t),\ldots, U_K(t),X(t) ) $. The Lyapunov function of the queue $Q(t)$ is $L( Q(t))= \frac{1}{2}[ \sum_{i=1}^K U_i^2(t) +X^2(t)            ]$. The Lyapunov drift of the queue $Q(t)$ is $L(Q(t+1))-L(Q(t))=\frac{1}{2}[ \sum_{i=1}^K U_i^2(t+1) +X^2(t+1)            ]-\frac{1}{2}[ \sum_{i=1}^K U_i^2(t) +X^2(t)] $, where
\begin{displaymath}
\begin{split}
U_i^2(t+1)- U_i^2(t) \leq&  U_i^2(t)+R_i^2(t)+\big( R_{si}(t, P(t), \varepsilon(t), \eta)I_i(t)\big)^2+\\
&2U_i(t)\big(R_i(t)-R_{si}(t, P(t), \varepsilon(t), \eta)I_i(t) \big) - U_i^2(t) \\
=&R_i^2(t)+\big( R_{si}(t, P(t), \varepsilon(t), \eta)I_i(t)\big)^2+2U_i(t)\big(R_i(t)-R_{si}(t, P(t), \varepsilon(t), \eta)I_i(t) \big),\\
X^2(t+1)-X^2(t)&\leq X^2(t)+P^2(t)+P_{av}^2+2X(t)\big(P(t)-P_{av}(t) \big) - X^2(t)\\
=&P^2(t)+P_{av}^2+2X(t)\big(P(t)-P_{av}(t) \big). \\
\end{split}
\end{displaymath}
The one time slot conditional Lyapunov drift of $Q(t)$ is $\Delta(Q(t))$,
\begin{displaymath}
\begin{split}
\Delta(Q(t))=&\mathbb{E} \{  L(Q(t+1))-L(Q(t))|   Q(t) \}\\
\leq&\mathbb{E} \{ \frac{1}{2}\sum_{i=1}^K [R_i^2(t)+\big( R_{si}(t, P(t), \varepsilon(t), \eta)I_i(t)\big)^2+2U_i(t)\big(R_i(t)-R_{si}(t, P(t), \varepsilon(t), \eta)I_i(t) \big)  ] +  \\    &\frac{1}{2}[P^2(t)+P_{av}^2+2X(t)\big(P(t)-P_{av}(t) \big) ]                      |   Q(t) \} \\
\leq& B + C + \mathbb{E}\{ \sum_{n=1}^K   U_i(t)\big(R_i(t)-R_{si}(t, P(t), \varepsilon(t), \eta)I_i(t) \big) |Q(t) \}
+\\
&\mathbb{E}\{ X(t)\big(P(t)-P_{av}(t) \big) |Q(t) \},
\end{split}
\end{displaymath}
Where $B=\frac{K A_{max}^2 + Rs_{max}^2}{2}$. Arrival data process $A_i(t)$ is bounded, so let $A_{max}$ be a constant that $A_{max}\geq A_i(t)\ \text{for all} \ i \ \text{and} \ t $.
In the real environment, the transmission power $P(t)$ is finite, and the channel gain $|h|^2$ is bounded by a sufficiently large constant $|h|^2_{max}$. Let $Rs_{max}$ be a constant that $Rs_{max} \geq R_{si}(t, P(t), \varepsilon(t), \eta)I_i(t)$  \text{for all}  $i$, $t$, $P(t)$, $\varepsilon(t)$, $\eta$, and channel states.
Thus, $B\geq\frac{1}{2}\sum_{i=1}^K [R_i^2(t)+\big( R_{si}(t, P(t), \varepsilon(t), \eta)I_i(t)\big)^2]$.
$C= \frac{P_{max}^2+P_{av}^2 }{2}$, $P_{max} \geq P(t)\ \text{for all} \ t$. Thus, $ C \geq\frac{1}{2}[P^2(t)+P_{av}^2 ]  $.

According to the stochastic optimization theory, the original optimization problem in Section \ref{formulation} can be solved by minimizing the
following drift-plus-penalty expression:
\begin{displaymath}
\begin{split}
\text{Minimize} \ & \Delta(Q(t))-V\mathbb{E } \{\sum_{i=1}^{K} \theta_i R_i(t)|Q(t)\}\\
 =&\mathbb{E} \{  L(Q(t+1))-L(Q(t))|   Q(t) \}-V\mathbb{E } \{\sum_{i=1}^{K} \theta_i R_i(t)|Q(t)\} \\
\leq & B + C + \mathbb{E}\{ \sum_{i=1}^K   U_i(t)\big(R_i(t)-R_{si}(t, P(t), \varepsilon(t), \eta)I_i(t) \big) |Q(t) \}
+\\
 &\mathbb{E}\{ X(t)\big(P(t)-P_{av} \big) |Q(t) \} -V\mathbb{E } \{\sum_{i=1}^{K} \theta_i R_i(t)|Q(t)\}\\
=&B+C- \mathbb{E} \{ X(t)P_{av}|Q(t)\} + \mathbb{E}\{ \sum_{i=1}^K   (U_i(t)-V\theta_i) R_i(t)  |Q(t) \}+\\
&\mathbb{E}\{ X(t)P(t)-  \sum_{i=1}^K U_i(t) R_{si}(t, P(t), \varepsilon(t), \eta)I_i(t)   |Q(t)      \}
\end{split}
\end{displaymath}

Thus, to minimize $\Delta(Q(t))-V\mathbb{E } \{\sum_{i=1}^{K} \theta_i R_i(t)|Q(t)\}$ is equal to minimizing $\sum_{i=1}^K   (U_i(t)-V\theta_i) R_i(t)$
 and $X(t)P(t)-  \sum_{i=1}^K U_i(t) R_{si}(t, P(t), \varepsilon(t), \eta)I_i(t)$ in every time slot $t$.

\subsection{Performance Analysis}
1. For user $i$, its data queue $U_{i}(t)$ is upper bounded by a constant $U_{i}^{max}$ for all $t$: $U_{i}(t)\leq U_{i}^{max}=V\theta_{i}+A_{max}$.

\textit{Proof}:
The proof is similar as in \cite{neely2006energy} \cite{neely2010stochastic}.
For all $i\in \{1,\ldots,K\}$,
when $t=0$, all queues are initialized to $0$. Thus, $U_i(0)\leq U_i^{max}$ is satisfied.
Assume that $U_i(t)\leq U_i^{max}$ for any time slot $t$. For time slot $t+1$, we need to consider two cases.
1) If $U_i(t)\leq U_i^{max}-A_{max}$, we have $U_{i}(t+1)\leq U_{i}(t)+R_i(t)\leq U_i^{max}-A_{max}+R_i(t)\leq U_i^{max}$.
2) If $U_i(t)\geq U_i^{max}-A_{max}$, then $ U_i(t)\geq V\theta_i +A_{max}-A_{max}=V\theta_i  $. Thus, according to the control scheme, no data is admitted, i.e., $U_i(t+1)\leq U_i^{max}$.

Thus, when parameter $V$ is small, the queue length is short (i.e., small queuing delay).
For queuing delay requirement $D_i$ of each user $i$ in the system, the parameter $V$ can be chosen as
$\text{Min}\displaystyle{_{i\in\{1,2,\ldots,K\}} (D_i-A_{max})/\theta_i}$.

2. The virtual power queue $X(t)$ is bounded by a constant $X_{max}$ for all $t$:
$X(t)\leq X_{max}=\gamma U_{max}+ P_{max} = \gamma V \theta_{max} + \gamma A_{max}+ P_{max}$,
where $U_{max}=\text{max}_{i\in\{1,2,\ldots,K\}}U_i^{max}$, $\theta_{max}=\text{max}_{i\in\{1,2,\ldots,K\}}\theta_i$,
and $\gamma$ is any constant that satisfies $R_{si}(t,  P(t), \varepsilon(t), \eta)\leq \gamma P(t) $ over all $i$, $t$, $P(t)$, $\varepsilon(t)$, $\eta$, and channel states.

\textit{Proof}:
The proof is similar as in \cite{neely2006energy} \cite{neely2010stochastic}.
There exists a finite $\gamma$ that $R_{si}(t,  P(t), \varepsilon(t), \eta)\leq \gamma P(t) $, e.g.,
$\gamma$ can be chosen as the maximum directional derivative of $C_{bi}(t,P(t),\varepsilon(t))$ with respect to $P(t)$, maximized over
all users, $\varepsilon(t)$ and channel states.
When $X(t)\geq \gamma U_{max} $, the following holds
\begin{displaymath}
\begin{split}
&\sum_{i=1}^K  U_i(t) R_{si}(t,  P(t), \varepsilon(t), \eta)I_i(t) - X(t)P(t) \\
\leq& \sum_{i=1}^K  U_i(t) R_{si}(t, P(t), \varepsilon(t), \eta)I_i(t) -  \gamma U_{max}P(t) \\
\leq&   \sum_{i=1}^K  U_i(t) \gamma P(t) I_i(t) -  \gamma U_{max}P(t)  \\
\leq& 0.
\end{split}
\end{displaymath}
The maximum $0$ is achieved when $P(t)=0$.
Therefore, when $X(t)\geq \gamma U_{max} $, $ X(t)$ will not further increase. Thus, $X(t)\leq \gamma U_{max}+ P_{max}= \gamma V \theta_{max} + \gamma A_{max}+ P_{max} $.

3.  The long term average admission rate achieved by our control scheme is within $(B+C)/V$ of the optimal value: $\lim\text{inf}_{T\rightarrow\infty}\frac{1}{T}\sum_{t=0}^{T-1}\sum_{i=1}^{K}\theta_i\mathbb{E}\{R_i(t) \}  \geq \sum_{i=1}^{K}\theta_i r_i^* -\frac{B+C}{V}$, where $B$ and $C$ are constants, and $\vec{r}^*=(r_1^*,\ldots,r_K^*)$ is the optimal admission rate vector.

\textit{Proof}:
The proof follows standard steps under the stochastic optimization framework \cite{neely2006energy} \cite{neely2010stochastic}.
The optimal admission rate vector $\vec{ r}^*=(r_1^*,\ldots,r_K^*)$ can in principle be achieved by the simple
backlog-independent admission control algorithm. Thus,
\begin{displaymath}
\begin{split}
&\mathbb{E} \{ R_i(t)|Q(t)  \} = \mathbb{E} \{ R_i(t)  \}= r_i^*, \\
&\mathbb{E} \{ R_{si}(t, P(t), \varepsilon(t), \eta)I_i(t)|Q(t)\}=\mathbb{E} \{ R_{si}(t,  P(t), \varepsilon(t), \eta)I_i(t)\}  \geq r_i^*, \\
&\mathbb{E} \{P(t)|Q(t)  \}= \mathbb{E} \{P(t) \} \leq P_{av}. \\
\end{split}
\end{displaymath}
Substitute three inequalities into the following right hand sides terms,
\begin{displaymath}
\begin{split}
&\text{Minimize} \Delta(Q(t))\\
 =&\mathbb{E} \{  L(Q(t+1))-L(Q(t))|   Q(t) \}+V\mathbb{E } \{\sum_{i=1}^{K} \theta_i R_i(t)|Q(t)\}-V\mathbb{E } \{\sum_{i=1}^{K} \theta_i R_i(t)|Q(t)\} \\
\leq& B + C+V\mathbb{E } \{\sum_{i=1}^{K} \theta_i R_i(t)|Q(t)\}+ \mathbb{E}\{ \sum_{i=1}^K   U_i(t)\big(R_i(t)-R_{si}(t,  P(t), \varepsilon(t), \eta)I_i(t) \big) |Q(t) \}+\\
&\mathbb{E}\{ X(t)\big(P(t)-P_{av} \big) |Q(t) \} -V\mathbb{E } \{\sum_{i=1}^{K} \theta_i R_i(t)|Q(t)\}\\
\leq& B+C +V\sum_{i=1}^{K}\theta_i\mathbb{E}\{R_i(t)|Q(t)\}-V\sum_{i=1}^{K}\theta_i r_i^*.
\end{split}
\end{displaymath}
Therefore,
\begin{displaymath}
\begin{split}
\mathbb{E}[\Delta(Q(t))]=&\mathbb{E}\{L\big(Q(t+1)\big)\}-\mathbb{E}\{L\big(Q(t)\big)\}\\
\leq & B+C +V\sum_{i=1}^{K}\theta_i\mathbb{E}\{R_i(t)\}-V\sum_{i=1}^{K}\theta_i r_i^*.
\end{split}
\end{displaymath}
Summing over $t=0,1,2,\ldots T-1$, we have
\begin{displaymath}
\begin{split}
\mathbb{E}\{L\big(Q(T)\big)\}-\mathbb{E}\{L\big(Q(0)\big)\}
\leq  T(B+C) + V   \sum_{t=0}^{T-1}\sum_{i=1}^{K}\theta_i\mathbb{E}\{R_i(t)\}-VT\sum_{i=1}^{K}\theta_i r_i^*.
\end{split}
\end{displaymath}
It follows that
\begin{displaymath}
\begin{split}
\frac{1}{T}\sum_{t=0}^{T-1}\sum_{i=1}^{K}\theta_i\mathbb{E}\{R_i(t) \}  \geq \sum_{i=1}^{K}\theta_i r_i^* -\frac{B+C}{V}-\mathbb{E}\{L\big(Q(0)\big)\}/TV
\end{split}
\end{displaymath}
Thus, $\lim\text{inf}_{T\rightarrow\infty}\frac{1}{T}\sum_{t=0}^{T-1}\sum_{i=1}^{K}\theta_i\mathbb{E}\{R_i(t) \}  \geq \sum_{i=1}^{K}\theta_i r_i^* -\frac{B+C}{V}$.

Thus, when $V$ becomes larger, the average admission rate is more close to the optimal value.

\section{Numerical Simulations}
In this section, we present the performance of our control scheme by simulations.
All the channels are Rayleigh fading. $\theta_i=1$ for all $i\in\{1,2,\ldots,K\}$.
The selection set $\Pi$ for total power $P(t)$ is $\{0, 100, 200, 300\}$, and the average power constraint $P_{av}$ is $200$.
The selection set $\Lambda$ for ratio $\varepsilon(t)$ is $\{0, \frac{1}{20}, \frac{2}{20}, \frac{3}{20},\ldots, \frac{19}{20},1\}$.

\subsection{Instantaneous Eavesdropping Channel Information}

In Fig. \ref{FullCSIadmissionrate} and Fig. \ref{FullCSIaveragequeuelength}, the system parameters for the simulation
are $N_A$=6, $N_E$=3, and $K$=2. Parameter $V$=5, 10, 20, and 100. The data arrival process for each user follows a binomial process with average $\lambda$, which varies from 1 to 30.
 Fig. \ref{FullCSIadmissionrate} shows that
1) For a fixed $V$, in the left part (i.e., the low arrival rate region), the average admission rate is equal to the arrival rate. The reason is that when the arrival rate is lower than the average secrecy channel capacity, all the arrival data are admitted into the queue. When the arrival rate is larger than the average secrecy channel capacity, the average admission rate is saturated with the increased arrival rate. For a fixed arrival rate, if parameter $V$ increases, the average admission rate is more close to the optimal value.
2) For fixed $V$ and $\lambda$, the average admission rate of the non-colluding case is higher than the one of the colluding case.
 Fig. \ref{FullCSIaveragequeuelength} shows that
  1) The average queue length is increased with the increment of parameter $V$ and arrival rate $\lambda$.
  2) For fixed $V$ and $\lambda$, the average queue length in the non-colluding case is shorter than the one of the colluding case.

\begin{figure*}[!t]
\centering
\subfigure[Average admission rate] { \label{FullCSIadmissionrate}
\includegraphics[width=7.5cm]{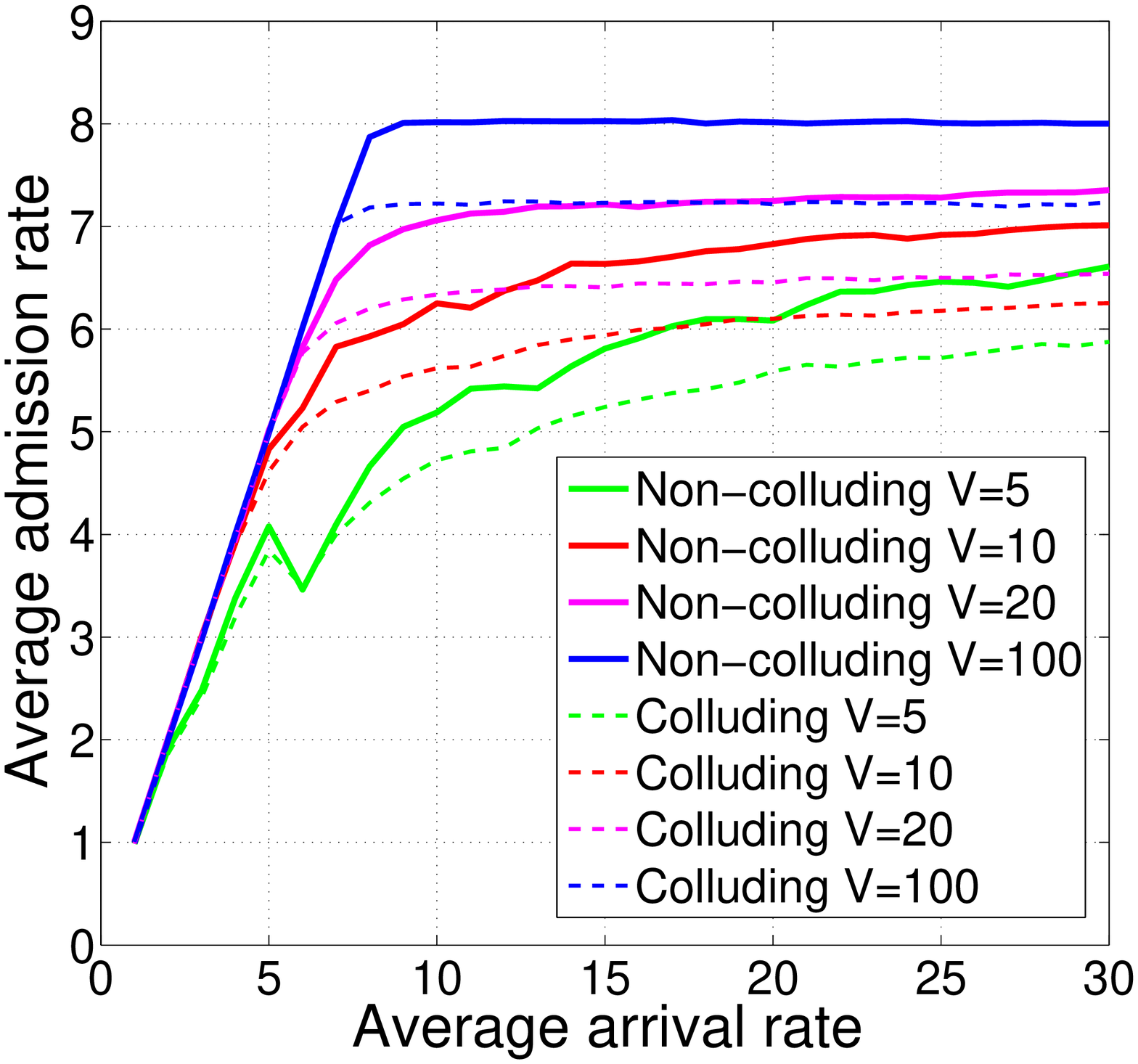}
}
\subfigure[Average queue length] { \label{FullCSIaveragequeuelength}
\includegraphics[width=7.8cm]{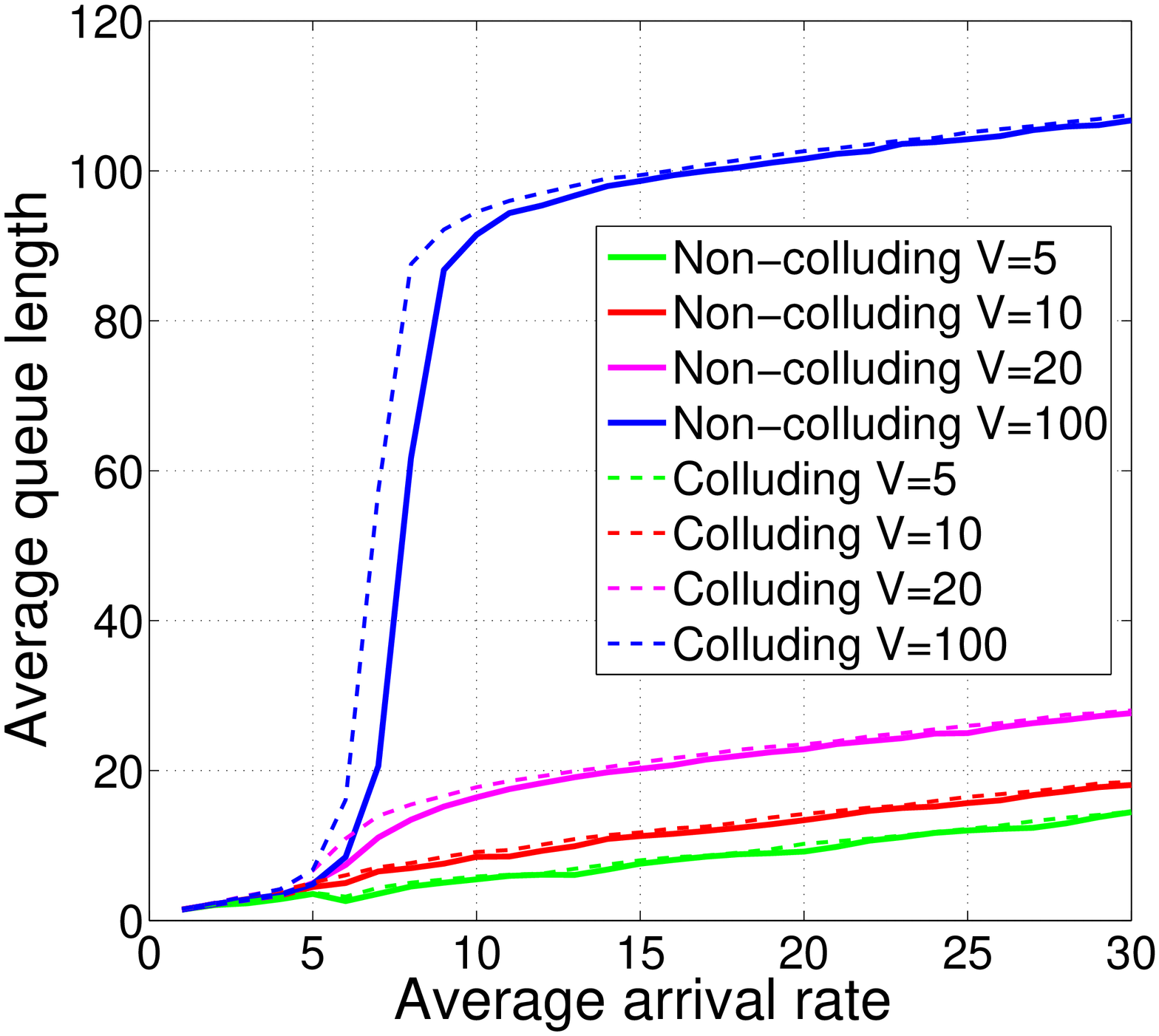}
}
\caption{In (a), average admission rate versus average arrival rate for both non-colluding and colluding cases in instantaneous eavesdropping channel information scenario. In (b), average queue length versus average arrival rate for both non-colluding and colluding cases in instantaneous eavesdropping channel information scenario.  }
\label{resource}

\end{figure*}

  In Fig. \ref{FullCSIantenadmissionrate} and Fig. \ref{FullCSIantenaveragequeuelength},
  the system parameters are $N_E$=3, $K$=2, and $V$=100. The data arrival process for each user follows a binomial process with average $\lambda=30$. The number of transmission antennas $N_A$=6, 8, 10, and 12.
  Fig. \ref{FullCSIantenadmissionrate} and Fig. \ref{FullCSIantenaveragequeuelength} show that
  1) As the number of antennas increases, the average admission rate is increased and the average queue length is correspondingly decreased for both the non-colluding and colluding cases.
  2) For the same number of antennas, the average admission rate of the non-colluding case is higher than the one of the colluding case, but the average queue length in the
  non-colluding case is shorter than the one in the colluding case.

\begin{figure*}[!t]
\centering
\subfigure[Average admission rate] { \label{FullCSIantenadmissionrate}
\includegraphics[width=7.55cm]{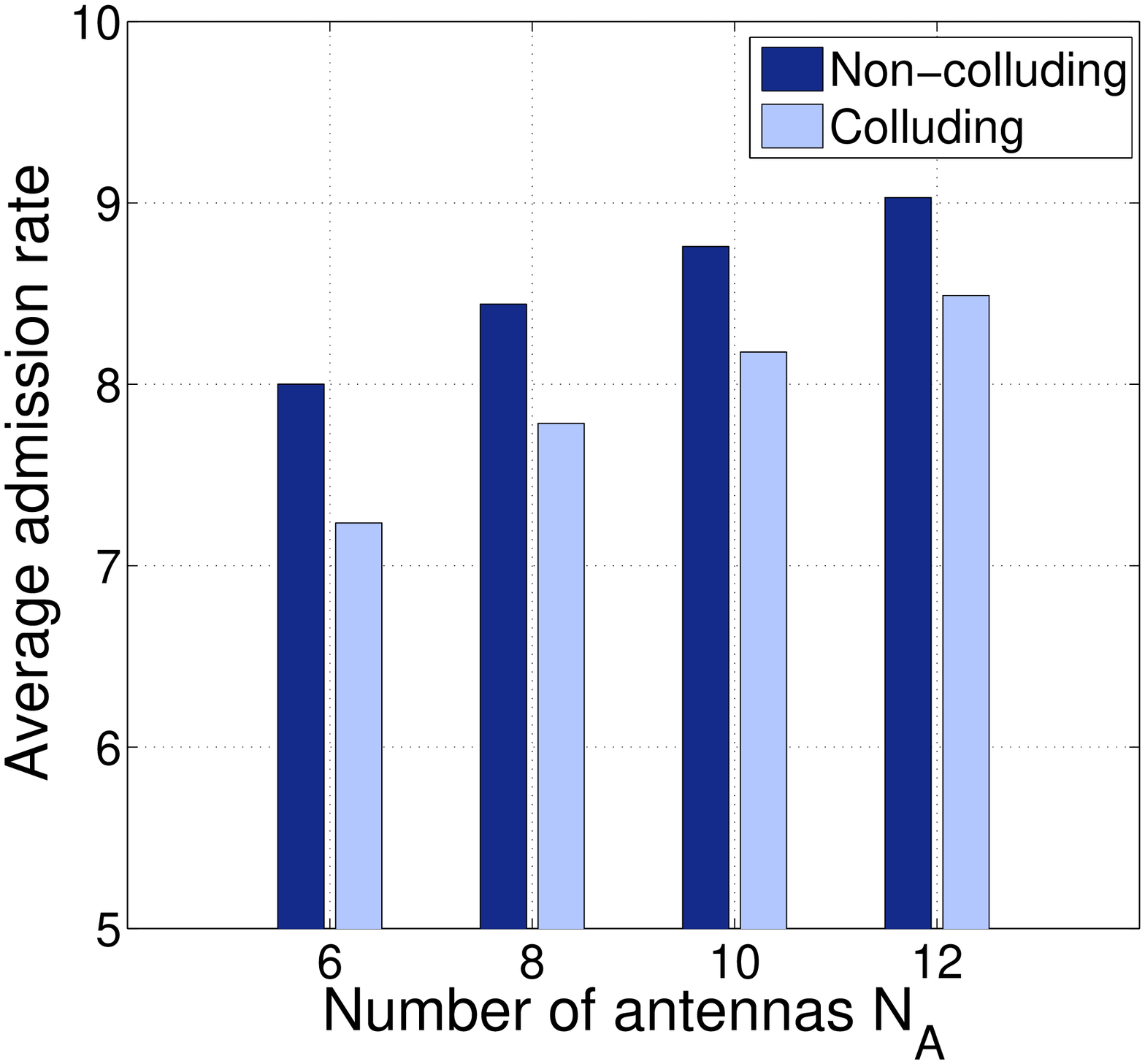}
}
\subfigure[Average queue length] { \label{FullCSIantenaveragequeuelength}
\includegraphics[width=7.6cm]{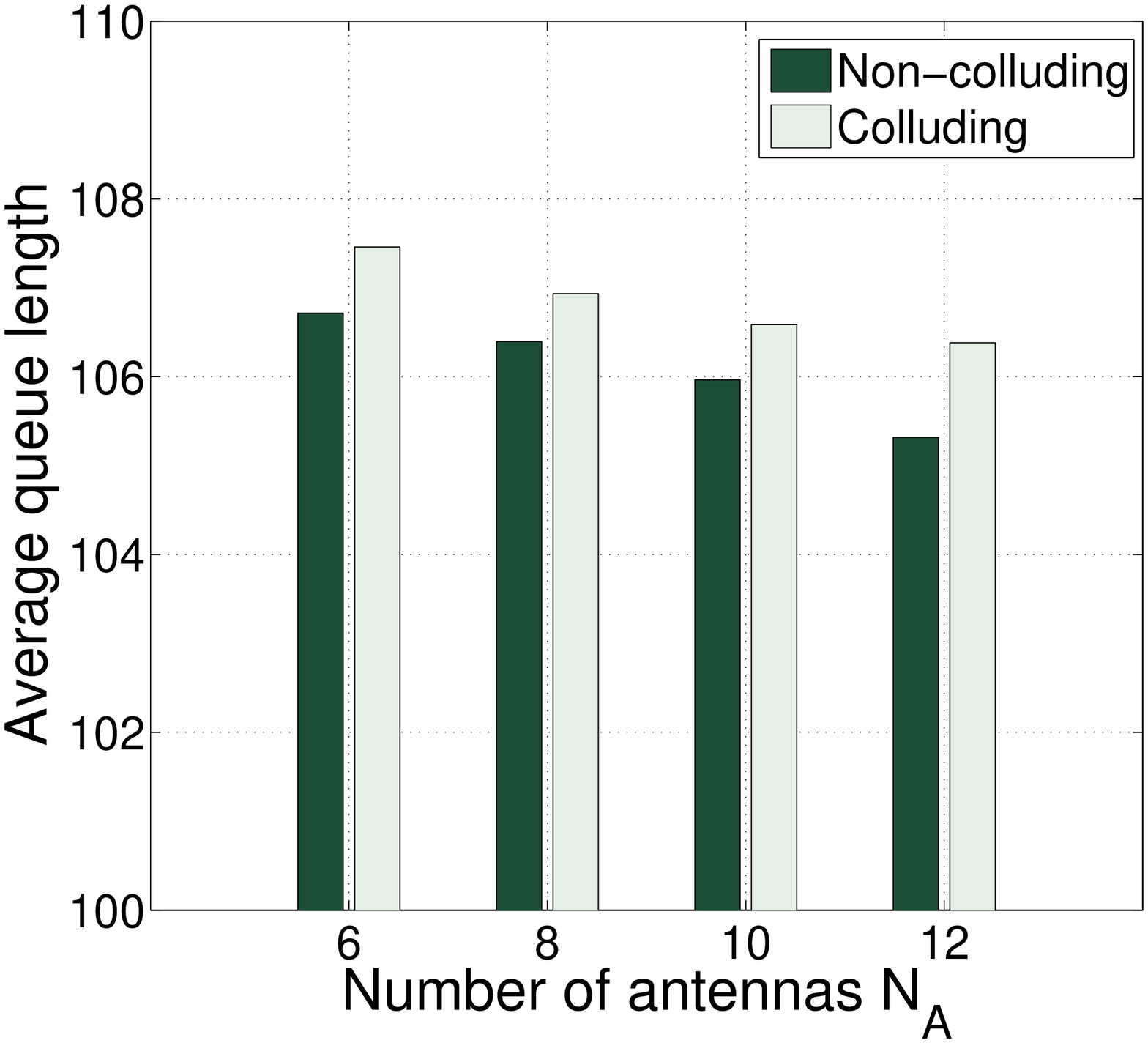}
}
\caption{In (a), average admission rate versus number of antennas for both non-colluding and colluding cases in instantaneous eavesdropping channel information scenario. In (b), average queue length versus number of antennas for both non-colluding and colluding cases in instantaneous eavesdropping channel information scenario. }
\label{resource}

\end{figure*}

\subsection{Partial Eavesdropping Channel Information}

In Fig \ref{PartialCSIadmissionrate} and Fig. \ref{PartialCSIaveragequeuelength}, the system parameters for the simulation are
$N_A$=6, $N_E$=3, $K$=2, and $\eta$=0.1. Parameter $V$=5, 10, 20, and 100. The data arrival process for each user follows a binomial process with average $\lambda$, which varies from 1 to 30.
Fig. \ref{PartialCSIadmissionrate} shows that:
1) For a fixed $V$, in the left part (i.e., the low arrival rate region), the average admission rate is equal to the arrival rate. The reason is that when the arrival rate is lower than the average secrecy channel capacity, all the arrival data are admitted into the queue. When the arrival rate is larger than the average secrecy channel capacity, the average admission rate is saturated with the increased arrival rate. For a fixed arrival rate, if parameter $V$ increases, the average admission rate is more close to the optimal value.
2) For fixed $V$ and $\lambda$, the average admission rate of the non-colluding case is higher than the one of the colluding case.
Fig. \ref{PartialCSIaveragequeuelength} shows that:
  1) The average queue length is increased with the increment of parameter $V$ and arrival rate $\lambda$.
  2) For fixed $V$ and $\lambda$, the average queue length in the non-colluding case is shorter than the one in the colluding case.

\begin{figure*}[!t]
\centering
\subfigure[Average admission rate] { \label{PartialCSIadmissionrate}
\includegraphics[width=7.35cm]{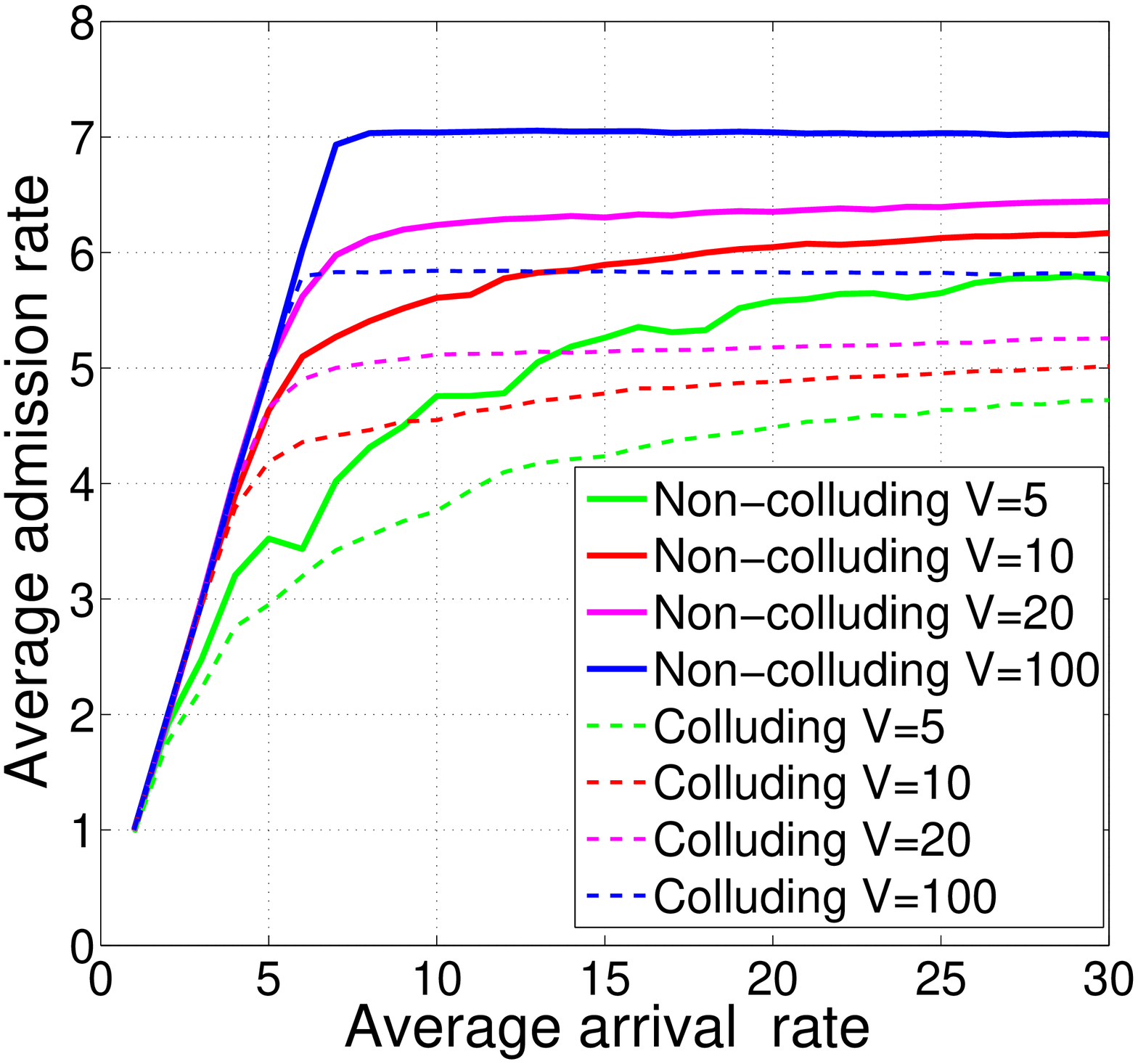}
}
\subfigure[Average queue length] { \label{PartialCSIaveragequeuelength}
\includegraphics[width=7.55cm]{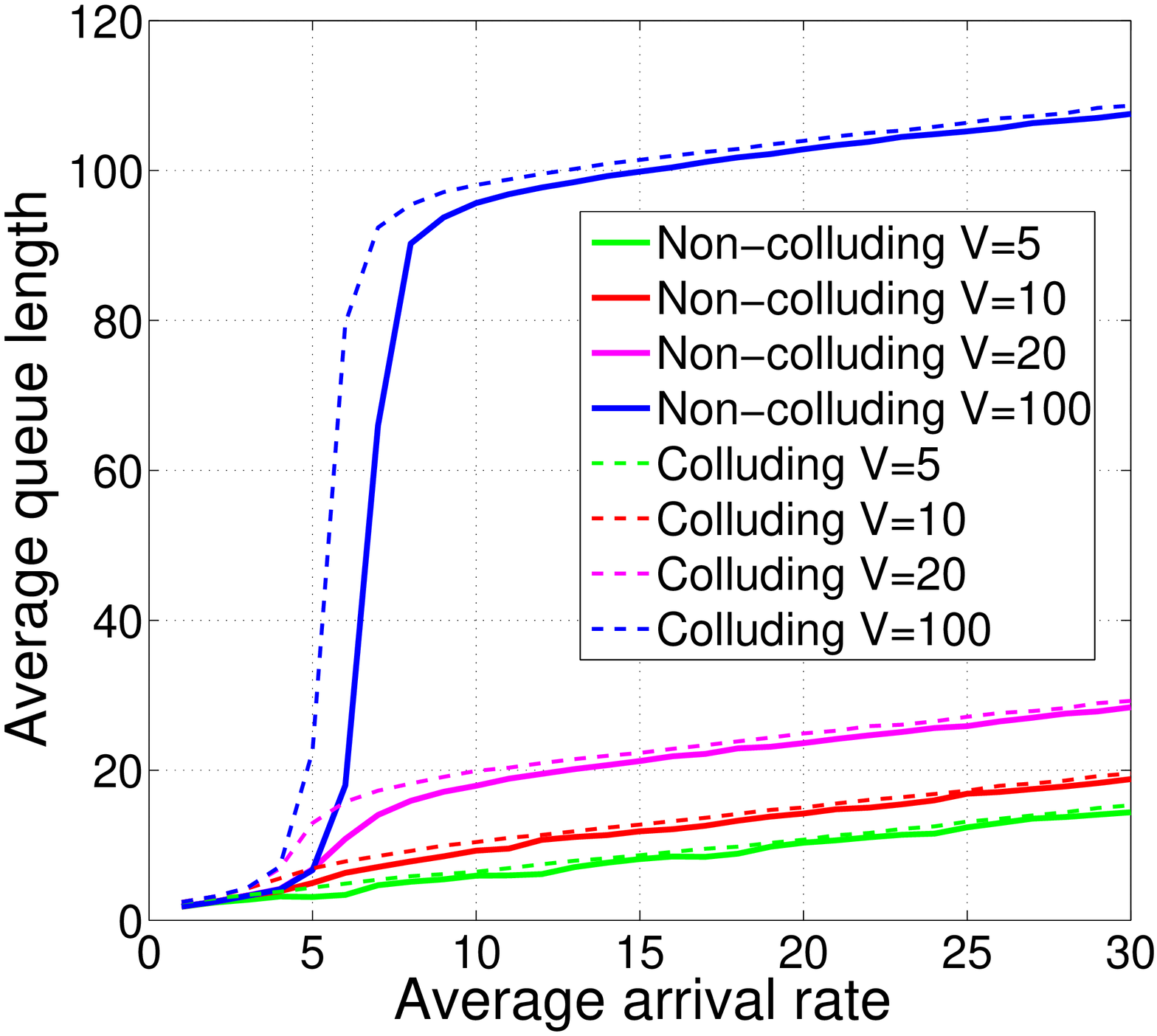}
}
\caption{ In (a), average admission rate versus average arrival rate for both non-colluding and colluding cases in partial eavesdropping channel information scenario. In (b), average queue length versus average arrival rate for both non-colluding and colluding cases in partial eavesdropping channel information scenario. }
\label{resource}

\end{figure*}

\begin{figure*}[!t]
\centering
\subfigure[Average admission rate] { \label{PartialCSIsecrecyoutageadmissionrate}
\includegraphics[width=7.5cm]{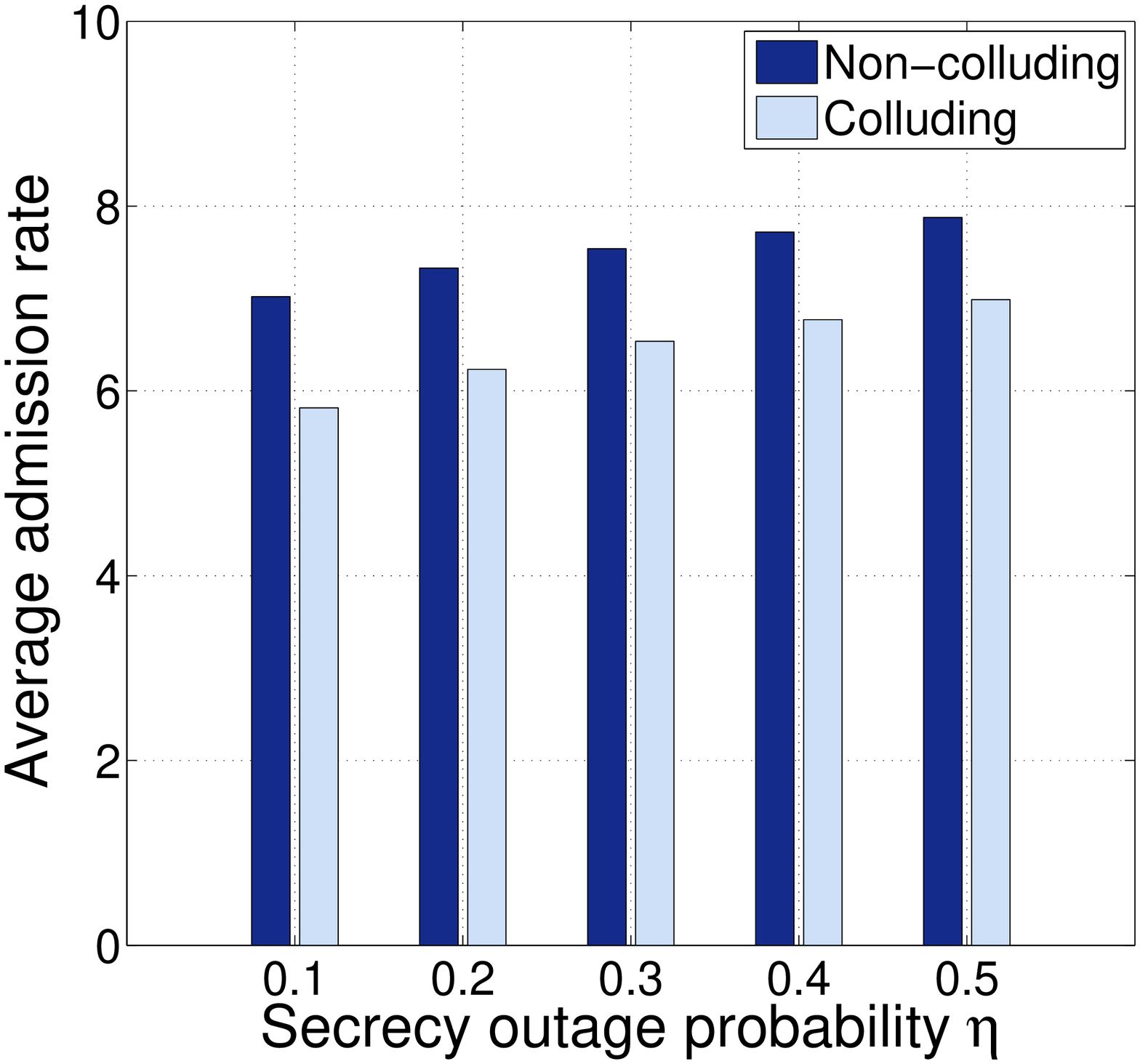}
}
\subfigure[Average queue length] { \label{PartialCSIsecrecyoutageaveragequeuelength}
\includegraphics[width=7.55cm]{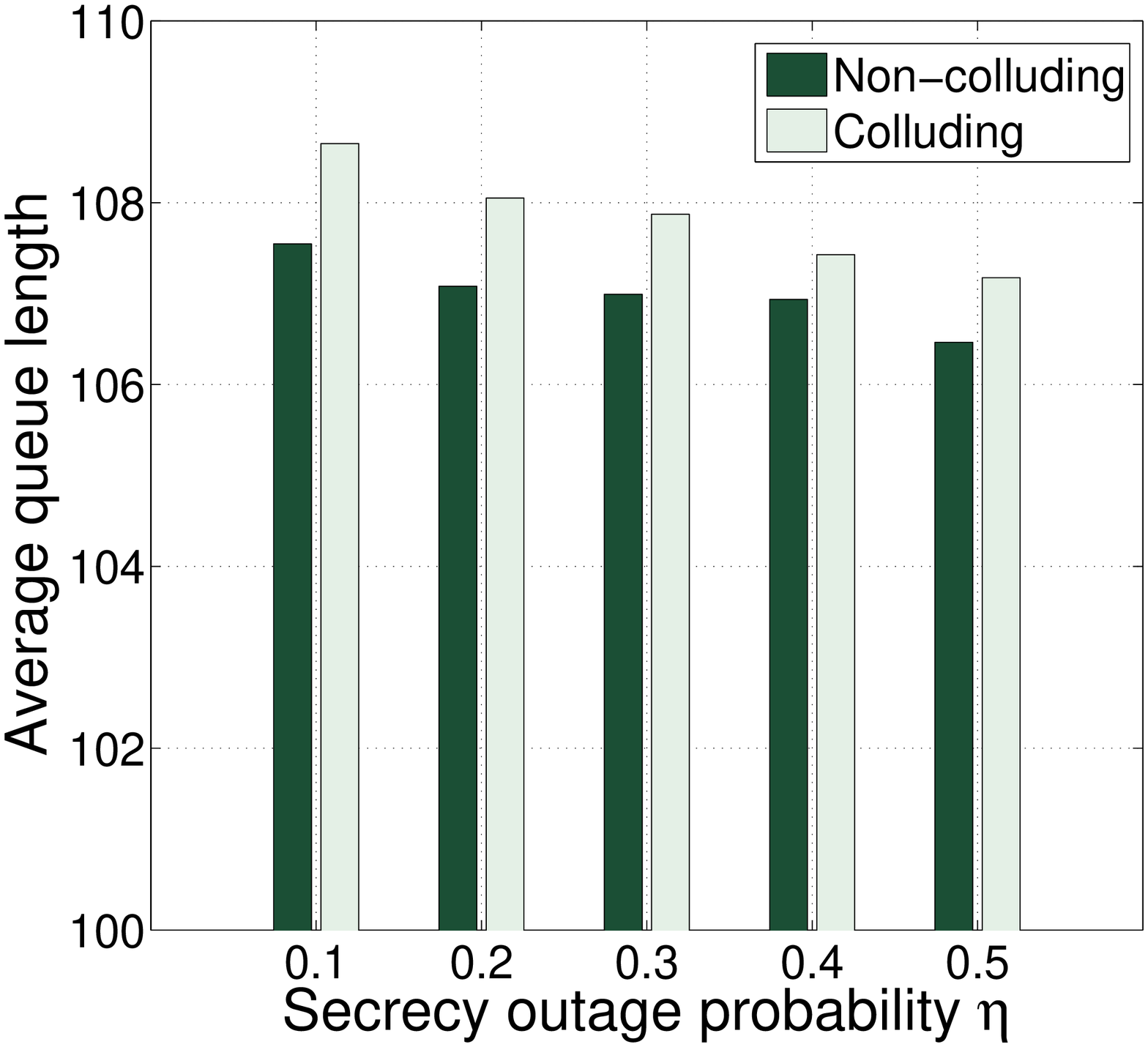}
}
\caption{ In (a), average admission rate versus secrecy outage probability for both non-colluding and colluding cases in partial eavesdropping channel information scenario. In (b), average queue length versus secrecy outage probability for both non-colluding and colluding cases in partial eavesdropping channel information scenario. }
\label{resource}

\end{figure*}

\begin{figure*}[!t]
\centering
\subfigure[Average admission rate] { \label{PartialCSIantenadmissionrate}
\includegraphics[width=7.5cm]{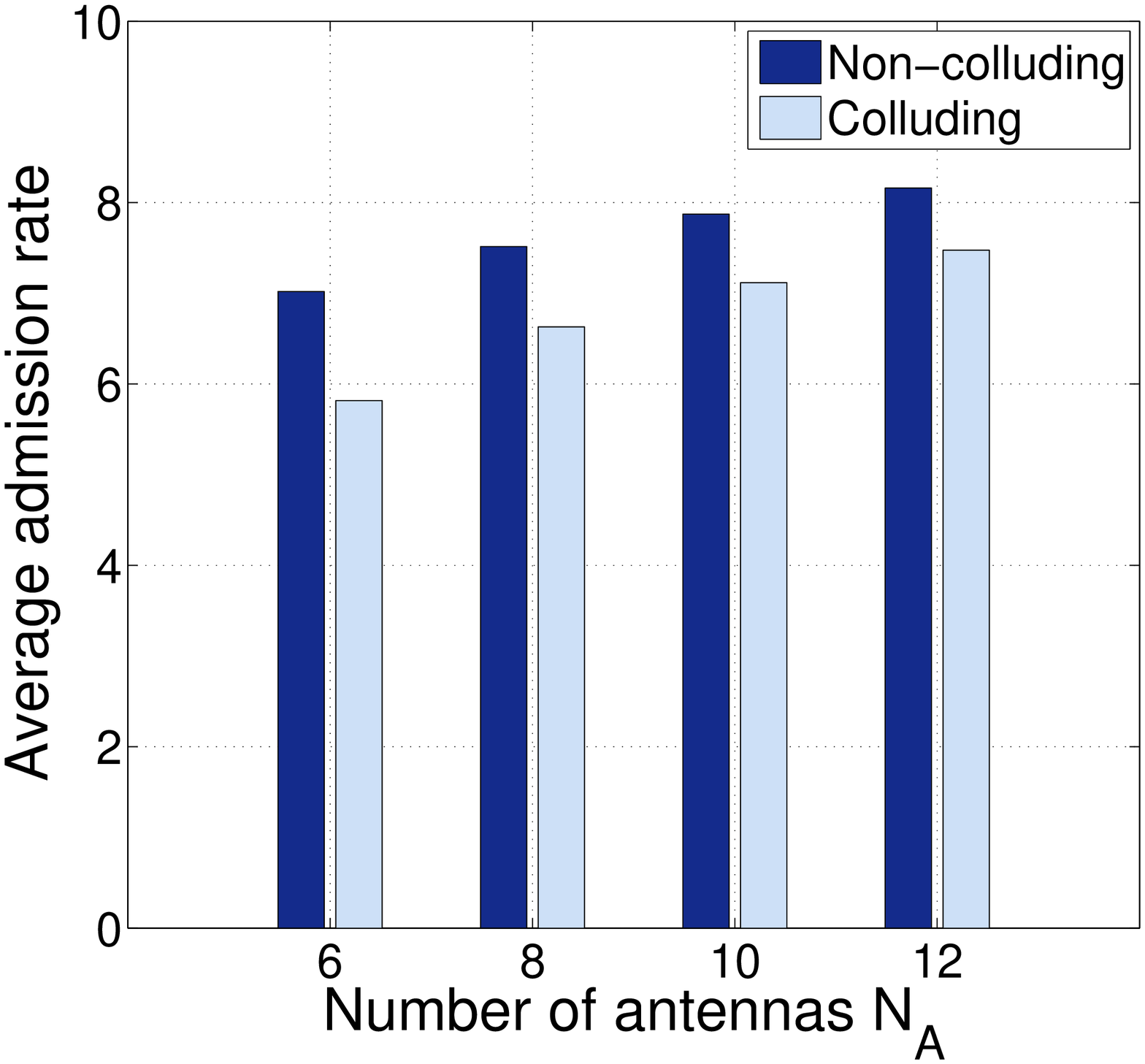}
}
\subfigure[Average queue length] { \label{PartialCSIantenaveragequeuelength}
\includegraphics[width=7.55cm]{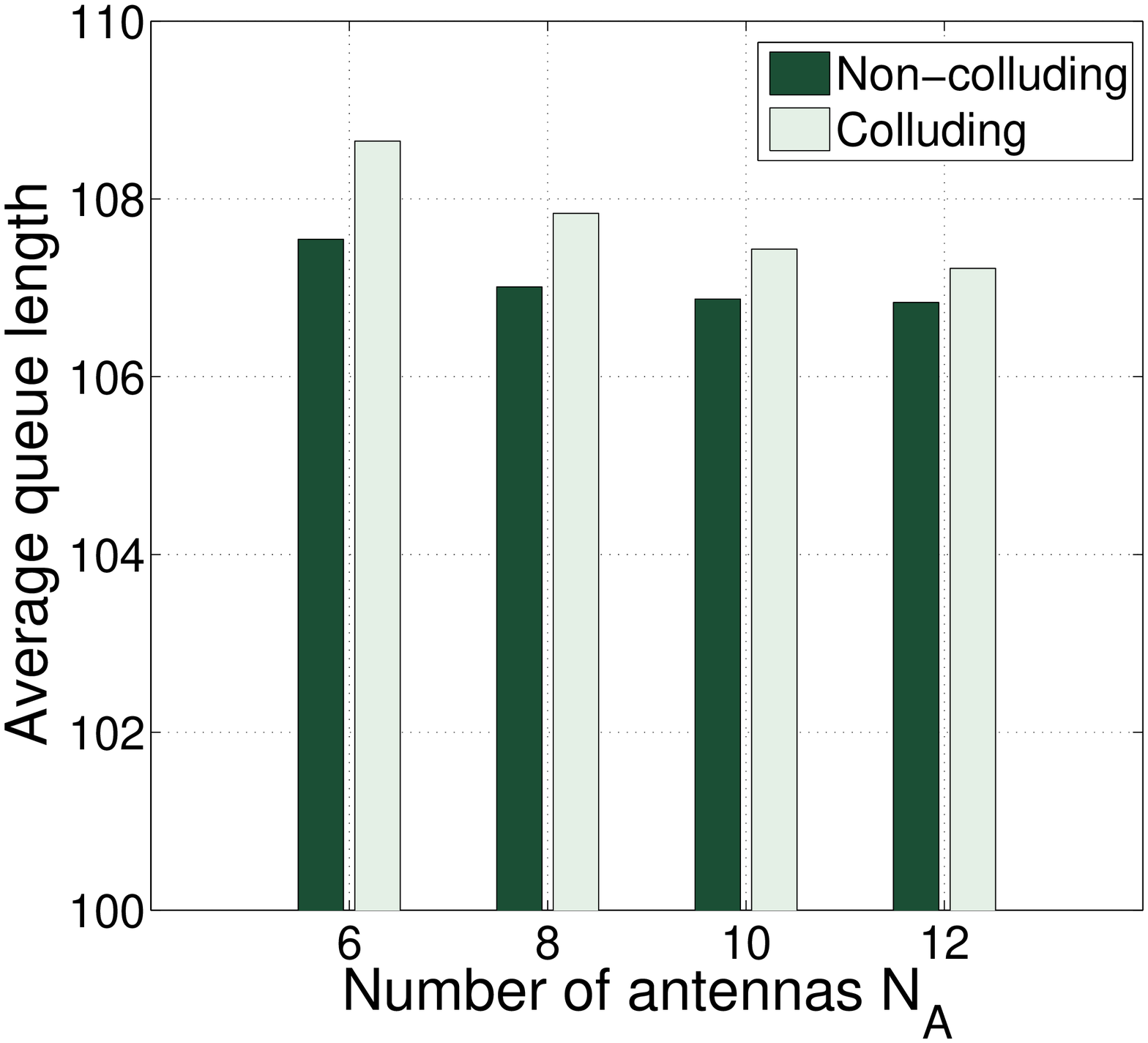}
}
\caption{ In (a), average admission rate versus number of antennas for both non-colluding and colluding cases in partial eavesdropping channel information scenario. In (b), average queue length versus number of antennas for both non-colluding and colluding cases in partial eavesdropping channel information scenario.  }
\label{resource}

\end{figure*}

  In Fig. \ref{PartialCSIsecrecyoutageadmissionrate} and Fig. \ref{PartialCSIsecrecyoutageaveragequeuelength},
  the system parameters are $N_A$=6, $N_E$=3, $K$=2, and $V$=100. The data arrival process for each user follows a binomial process with average $\lambda=30$.
  The secrecy outage probability $\eta$=0.1, 0.2, 0.3, 0.4 and 0.5.
  Fig. \ref{PartialCSIsecrecyoutageadmissionrate} and Fig. \ref{PartialCSIsecrecyoutageaveragequeuelength} show that
  1) For both non-colluding and colluding cases, when the secrecy requirement is loose (i.e., the secrecy outage probability $\eta$ becomes larger), the average admission rate is increased and the average queue length is correspondingly reduced.
  2) For the same secrecy outage $\eta$, the average admission rate of the non-colluding case is higher than the one of the colluding case, but the average queue length in the
  non-colluding case is shorter than the one in the colluding case.

  In Fig. \ref{PartialCSIantenadmissionrate} and Fig. \ref{PartialCSIantenaveragequeuelength},
  the system parameters are $N_E$=3, $K$=2, $V$=100, and $\eta$=0.1. The data arrival process for each user follows a binomial process with average $\lambda=30$.
  The number of transmission antennas $N_A$=6, 8, 10, and 12.
  Fig. \ref{PartialCSIantenadmissionrate} and Fig. \ref{PartialCSIantenaveragequeuelength} show that
  1) As the number of antennas increases, the average admission rate is increased and the average queue length is correspondingly decreased for both the non-colluding and colluding cases.
  2) For the same number of antennas, the average admission rate of the non-colluding case is higher than the one of the colluding case, but the average queue length in the
  non-colluding case is shorter than the one in the colluding case.

\section{Conclusion}

In this paper, we considered the cross-layer resource allocation problem for the multi-user secure communication system in both the sender having instantaneous and partial eavesdropping channel information scenarios.
In each scenario, for both non-colluding and colluding eavesdropping cases, we designed admission controller based on the information in the upper layer and power controller with the information from physical layer and upper layer.
Simulation results validate the effectiveness of our scheme.


\end{abstract}


%
\IEEEpeerreviewmaketitle

\ifCLASSOPTIONcaptionsoff
  \newpage
\fi



%
%



\bibliographystyle{unsrt}
\bibliography{IEEEabrv}

\begin{thebibliography}{10}

\bibitem{shannon1949communication}
C.E. Shannon.
\newblock Communication theory of secrecy systems.
\newblock {\em Bell System Technical Journal}, 28(4):656--715, 1949.

\bibitem{wyner1975wire}
A.D. Wyner.
\newblock The wire-tap channel.
\newblock {\em Bell System Technical Journal}, 54(8):1355--1387, 1975.

\bibitem{leung1978gaussian}
S.~Leung-Yan-Cheong and M.~Hellman.
\newblock The gaussian wire-tap channel.
\newblock {\em IEEE Transactions on Information Theory}, 24(4):451--456, 1978.

\bibitem{csiszar1978broadcast}
I.~Csisz{\'a}r and J.~Korner.
\newblock Broadcast channels with confidential messages.
\newblock {\em IEEE Transactions on Information Theory}, 24(3):339--348, 1978.

\bibitem{bloch2006wireless}
M.~Bloch, J.~Barros, M.R.D. Rodrigues, and S.W. McLaughlin.
\newblock Wireless information-theoretic security.
\newblock {\em IEEE Transactions on Information Theory}, 54(6):2515 --2534,
  2008.

\bibitem{gopala2008secrecy}
P.K. Gopala, L.~Lai, and H.~El~Gamal.
\newblock On the secrecy capacity of fading channels.
\newblock {\em IEEE Transactions on Information Theory}, 54(10):4687--4698,
  2008.

\bibitem{liang2008secure}
Y.~Liang, H.V. Poor, and S.~Shamai.
\newblock Secure communication over fading channels.
\newblock {\em IEEE Transactions on Information Theory}, 54(6):2470--2492,
  2008.

\bibitem{shafiee2007achievable}
S.~Shafiee and S.~Ulukus.
\newblock Achievable rates in gaussian miso channels with secrecy constraints.
\newblock In {\em Proc. IEEE International Symposium on Information Theory
  (ISIT)}, pages 2466--2470, 2007.

\bibitem{huang2011robust}
J.~Huang and A.~Swindlehurst.
\newblock Robust secure transmission in miso channels based on worst-case
  optimization.
\newblock {\em IEEE Transactions on Signal Processing}, 60(4):1696--1707, 2011.

\bibitem{li2011optimal}
Q.~Li and W.~Ma.
\newblock Optimal and robust transmit designs for miso channel secrecy by
  semidefinite programming.
\newblock {\em IEEE Transactions on Signal Processing}, 59(8):3799 --3812,
  2011.

\bibitem{goel2008guaranteeing}
S.~Goel and R.~Negi.
\newblock Guaranteeing secrecy using artificial noise.
\newblock {\em IEEE Transactions on Wireless Communications}, 7(6):2180--2189,
  2008.

\bibitem{romero2011physical}
N.~Romero-Zurita, M.~Ghogho, and D.~McLernon.
\newblock Physical layer security of mimo frequency selective channels by
  beamforming and noise generation.
\newblock In {\em Proc. European Signal Processing Conference (EUSIPCO)}, pages
  829 --833, 2011.

\bibitem{6094170}
N.~Romero-Zurita, M.~Ghogho, and D.~McLernon.
\newblock Outage probability based power distribution between data and
  artificial noise for physical layer security.
\newblock {\em IEEE Signal Processing Letters}, 19(2):71 --74, 2012.

\bibitem{zhou2010secure}
X.~Zhou and M.R. McKay.
\newblock Secure transmission with artificial noise over fading channels:
  achievable rate and optimal power allocation.
\newblock {\em IEEE Transactions on Vehicular Technology}, 59(8):3831--3842,
  2010.

\bibitem{zhou2011rethinking}
X.~Zhou, M.R. McKay, B.~Maham, and A.~Hjorungnes.
\newblock Rethinking the secrecy outage formulation: A secure transmission
  design perspective.
\newblock {\em IEEE Communications Letters}, 15(3):302--304, 2011.

\bibitem{zhangbenefits}
X.~Zhang, X.~Zhou, and M.R. McKay.
\newblock Benefits of multiple transmit antennas in secure communication: A
  secrecy outage viewpoint.
\newblock In {\em Proc. IEEE ASILOMAR}, pages 212--216, 2011.

\bibitem{eryilmaz2006joint}
A.~Eryilmaz and R.~Srikant.
\newblock Joint congestion control, routing, and mac for stability and fairness
  in wireless networks.
\newblock {\em IEEE Journal on Selected Areas in Communications},
  24(8):1514--1524, 2006.

\bibitem{neely2006energy}
M.J. Neely.
\newblock Energy optimal control for time-varying wireless networks.
\newblock {\em IEEE Transactions on Information Theory}, 52(7):2915--2934,
  2006.

\bibitem{neely2010stochastic}
M.J. Neely.
\newblock Stochastic network optimization with application to communication and
  queueing systems.
\newblock {\em Synthesis Lectures on Communication Networks}, 3(1):1--211,
  2010.

\bibitem{koksal2010control}
C.E. Koksal, O.~Ercetin, and Y.~Sarikaya.
\newblock Control of wireless networks with secrecy.
\newblock In {\em Proc. IEEE ASILOMAR}, pages 47--51, 2010.

\bibitem{mao2011towards}
Z.~Mao, C.E. Koksal, and N.B. Shroff.
\newblock Towards achieving full secrecy rate in wireless networks: A control
  theoretic approach.
\newblock In {\em Proc. IEEE Information Theory and Applications Workshop
  (ITA)}, pages 1--8, 2011.

\bibitem{liang2008wireless}
Y.~Liang, H.V. Poor, and L.~Ying.
\newblock Wireless broadcast networks: reliability, security, and stability.
\newblock In {\em Proc. IEEE Information Theory and Applications Workshop
  (ITA)}, pages 249--255, 2008.

\bibitem{ioannis2011feedback}
K.~Ioannis, T.~John~S, M.L. Steve, G.~Peter~M,
\newblock A feedback-based transmission for wireless networks with energy and
  secrecy constraints.
\newblock {\em EURASIP Journal on Wireless Communications and Networking},
  2011, 2011.

\end{thebibliography}

%




\end{document}